\documentclass[pdflatex,sn-mathphys]{sn-jnl}
\usepackage{amsmath,amsthm,mathtools}
\usepackage{amssymb}
\usepackage[T1]{fontenc}
\usepackage{lmodern}
\usepackage[utf8]{inputenc}
\usepackage[english]{babel}
\usepackage{upgreek}
\usepackage{xcolor}
\usepackage{soul}
\UseRawInputEncoding

\usepackage{lineno}

\begin{document}

\title{Optically driven spin precession in polariton condensates}
\author*[1]{\fnm{Ivan} \sur{Gnusov}}\email{ivan.gnusov@skoltech.ru}

\author[1]{\fnm{Stepan} \sur{Baryshev}}

\author[2,3]{\fnm{Helgi} \sur{Sigur{\dh}sson}}

\author[1]{\fnm{Kirill} \sur{Sitnik}}

\author[1]{\fnm{Julian} \sur{T\"{o}pfer}}

\author[1]{\fnm{Sergey} \sur{Alyatkin}}

\author*[1,3]{\fnm{Pavlos G.} \sur{Lagoudakis}}\email{Pavlos.Lagoudakis@soton.ac.uk}

\affil[1]{\orgname{Hybrid Photonics Laboratory, Skolkovo Institute of Science and Technology}, \orgaddress{\street{Territory of Innovation Center Skolkovo, Bolshoy Boulevard 30, building 1}, \postcode{121205} \city{Moscow}, \country{Russia}}}

\affil[2]{Science Institute, University of Iceland, Dunhagi 3, IS-107, Reykjavik, Iceland}

\affil[3]{\orgdiv{School of Physics and Astronomy}, \orgname{ University of Southampton}, \orgaddress{\city{Southampton}, \postcode{SO17 1BJ}, \country{UK}}}

\abstract{External driving of spinor degrees of freedom by magnetic or optical fields in  quantum systems underpin many applications ranging from nuclear magnetic resonance to coherent state control in quantum computing. Although spinor polariton condensates are offering a flexible platform for spinoptronic applications, strong inter-particle interactions limit their spin coherence. Here, we introduce an all-optically driven spin precession in microcavity polariton condensates that eliminates depolarisation, through a radio frequency modulation of a spatially rotating, asymmetric exciton reservoir that both confines, and actively replenishes the polariton condensate. We realise several GHz driven spin precession with a macroscopic spin coherence time that is limited only by the extraneous to the condensate, frequency drift of the composite pumping sources. Our observations are supported by mean field modelling and evidence a driven-dissipative quantum fluidic analogue of the nuclear magnetic resonance effect.}

\maketitle


The spin of particles is a quantum degree of freedom defining their
behavior in magnetic fields. Understanding and controlling the spin properties
of materials opens a route to numerous applications such as quantum computing~\cite{Xin_ChiPhysB2018}, spintronic solid state devices~\cite{Yang_NatRevPhys2021} and high resolution imaging and spectroscopy~\cite{emsley2013high}. An effect, common to all spins in transverse magnetic fields, is the Larmor precession wherein spins precess steadily around the magnetic field lines. This phenomenon has been observed in various magnetic systems~\cite{spin_book2001, nmr_rabi, Broholm_Science2020}, and is currently used in commercial applications such as nuclear magnetic resonance (NMR) spectroscopy and magnetic resonance imaging to probe the composition of biological tissues and materials~\cite{sanders1988modern}. Moreover, all-optical manipulation of the spinor degrees of freedom is utilised to drive spin precession in dilute atomic gases~\cite{opt_driv_PhysRevLett.6.280} and Bose-Einstein condensates~\cite{PhysRevLett.95.050401_larmor_bose} with application in high-precision magnetometry~\cite{gas_magnetometer_nature}. Recently, the presence of intrinsic energy splittings in microcavity polaritons was shown to act as an effective magnetic field that results in self-induced Larmor precession of spinor polariton condensates under non-resonant elliptically polarised optical excitation~\cite{baryshev_prl, coher_revivals}.

Exciton-polaritons (or just polaritons) are composite quasiparticles formed in the strong coupling regime between photons and excitons in semiconductor microcavities~\cite{kavokin_microcavities_2007}. 
A polariton condensate is characterised by a spinor degree of freedom that results in the $\sigma^\pm$ circular polarisation of the emitted cavity light. The strong non-linearities, evidenced even at the few photons level~\cite{Zasedatelev2021nature, Kuriakose_NatPho2022}, renders them a promising candidate for  spinoptronic devices~\cite{Amo_NatPho2010, Cerna_NatComm2013, askitopoulos_all-optical_2018, pol_review, Liew_PhysE2011}. A plethora of nonlinear polariton spin effects have been reported including bistability~\cite{pickup_optical_2018, Ohadi19, Sigurdsson_PRR2020}, spin bifurcations~\cite{ohadi_spontaneous_2015}, the optical spin Hall effect~\cite{leyder_observation_2007}, half-quantised vortices~\cite{Lagoudakis2009_halfquantised} and recent theoretical proposal of rotation induced spin resonance~\cite{Yulin_arxiv2022}.
The manipulation of the spinor of polariton condensates, beyond the impractical application of external magnetic fields~\cite{Walker_PRL2011} can be realised by controlling the polarisation of the optical pump and therefore the spin imbalance of the injected carriers~\cite{gnusov_prb,pickup_optical_2018, baryshev_prl, coher_revivals}. 
However, the ultrafast dynamics of polariton condensates has limited the implementation of optically driven spin precession, analogously to atomic Bose-Einstein condensates~\cite{opt_driv_PhysRevLett.6.280}. 

In this article, we report on an optically driven spin precession in polariton condensates. We demonstrate controllable driven GHz rotation of the condensate pseudospin in the equatorial plane of the Bloch (Poincar\'{e}) sphere or, equivalently, the GHz rotation of linear polarisation of the cavity emitted light. We utilise the beat note of two frequency-detuned and spatially structured laser beams to create a rotating optical trap of broken axial symmetry which confines and sustains the condensate with particles. Upon condensation the polariton pseudospin spontaneously aligns itself along the short axis of the trap in the cavity plane~\cite{gnusov_prapl} and rotates in-step with it. At GHz rotation frequencies we observe a resonant signature in the spin coherence with dramatic increase of spin coherence time. This striking behaviour appears when the driving frequency is in resonance with the condensate's self-induced Larmor precession frequency. Our work underpins the ultrafast dynamical response of polariton condensates and the exciton reservoir under GHz driving with unprecedented persistence to retain cyclical dynamics, reminiscent of optically driven dilute alkaline gases~\cite{opt_driv_PhysRevLett.6.280}. 

\section*{Results}

The polariton condensate can be assigned to a spinor order parameter $\Psi = (\psi_+,\psi_-)^\text{T}$ with a normalised pseudospin defined
\begin{equation}
\mathbf{S} = \frac{\langle \Psi \lvert \hat{\boldsymbol{\sigma}} \rvert \Psi \rangle}{\langle \Psi \vert \Psi \rangle}
\end{equation}
where $\hat{\boldsymbol{\sigma}}$ is the Pauli matrix vector. The components of the pseudospin correspond to the Stokes parameters of the emitted cavity light. Therefore, when polaritons spontaneously decay, the emitted photons carry all the information of the condensate including the spin. This salient feature permits polarimetry measurements to characterise the spin state of the polariton fluid, which is not so easy in electronic or cold atoms systems.
 
\begin{figure}[t]
    \centering
    \includegraphics[width=0.6\columnwidth]{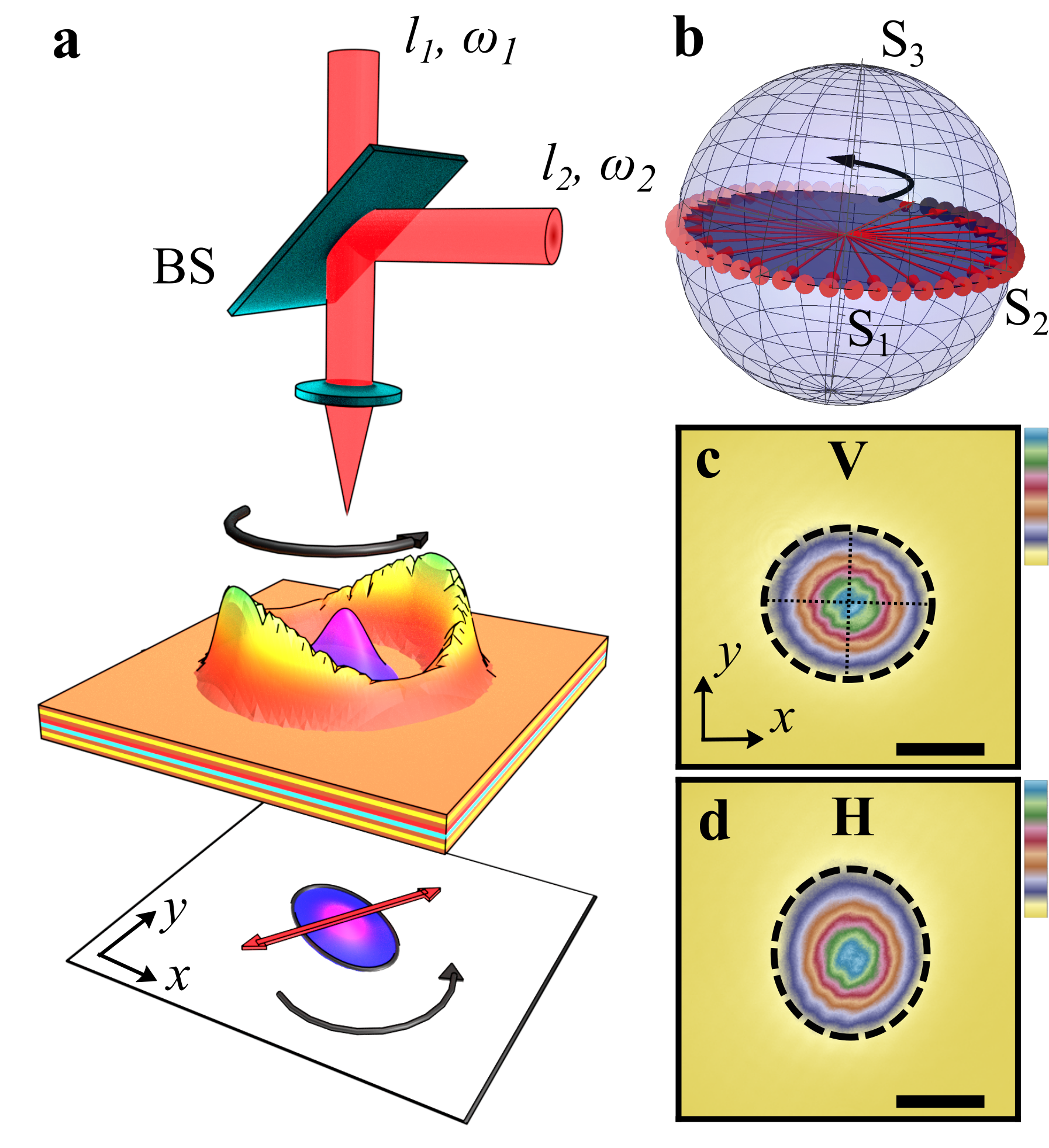}
    \caption{\textbf{Schematic of the excitation part of experimental setup and polariton condensate intensity distribution.} \textbf{a} The schematic of experimental setup demonstrating two frequency detuned ($f_{1,2} = \omega_{1,2}/2\pi$) perfect vortex beams overlapped on the beamsplitter(BS). The resultant rotating non-resonant excitation pattern (colored in red to yellow gradient) is projected onto the microcavity sample to form the rotating elliptical polariton condensate (colored in blue to purple gradient). The bottom of the panel \textbf{a} depicts the planar spinor projection of the condensate. \textbf{b} Polarisation rotation in the equatorial plane of the Poincar\'{e} sphere. 
    Normalised time-averaged intensities of the \textbf{c} vertical and \textbf{d} horizontal polarisation projections of the condensate emission (the black dashed lines are to guide the eye). The black scale bar corresponds to 5 $\upmu$m. }
    \label{fig1}
\end{figure}

In our experiment, polaritons are excited in an inorganic $2 \lambda$ GaAs/AlAs$_{0.98}$P$_{0.02}$ microcavity with embedded InGaAs quantum wells, held in cryostat at $4$ K~\cite{cilibrizzi_polariton_2014}. The rotating excitation pattern is created with the similar technique as in~\cite{rotatingbucket}. Polaritons are injected non-resonantly, with two continuous wave lasers detuned by $\approx 100$ meV higher from the polariton resonance. 
These excitation lasers are stabilised at frequencies $\omega_1$ and $\omega_2$, correspondingly. 
Each laser beam profile is shaped independently with a spatial light modulator in a form of "perfect vortex"~\cite{Chen2013_perfectvortex}, implying an annulus of equal diameter, but of opposite corresponding orbital angular momenta $l_{1,2}=\pm1$ (see Supplementary Information section S1). The annular beams are overlapped on the beamsplitter (see Figure~\ref{fig1}\textbf{a}). The resultant in-plane beating note profile rotates  with frequency $\omega =2\pi f= (\omega_1-\omega_2)/(l_1-l_2)$~\cite{rotatingbucket, rotating}. In order to create an elliptically elongated excitation pattern, as in Figure~\ref{fig1}\textbf{a}, we make the intensity of one of the beams to be  $10\% \to 20\%$ (depending on rotation speed) to that of the other.

When the excitation reaches threshold power ($P_{th}$), polaritons undergo a transition into a macroscopically occupied coherent state - a polariton condensate. The condensation occurs into the trap ground state with some finite ellipticity in real space, dictated by the effective elongation of the optical trap~\cite{gnusov_prapl}.
The broken axial symmetry of the trap, alongside with the inherent cavity TE-TM splitting~\cite{Panzarini_PhySolSt1999}, causes fine-structure splitting of the trap ground state resulting in a condensate with a definite pseudospin orientation (polarisation) along the short axis of the optical trap~\cite{gnusov_prapl, Rubo_arxiv2022}. This can be interpreted as an effective magnetic field coming from the confinement acting on the pseudospin written,
\begin{equation} \label{eq.Om}
\boldsymbol{\Omega}_\text{trap} = \begin{pmatrix} \Omega_\parallel \cos{[2 \theta_\text{small}(t)]} \\ \Omega_\parallel \sin{[2 \theta_\text{small}(t)]} \\ 0
\end{pmatrix},
\end{equation}
where $\theta_\text{small}$ is the angle of the trap small axis in the cavity plane and $\Omega_\parallel$ is proportional to the spatial ellipticity of the trap~\cite{gnusov_prapl}. The rotating excitation profile described above implies $\theta_\text{small} = \omega t$ meaning we have designed a time-periodic effective magnetic field rotating in the cavity plane. The condensate pseudospin will then follow this rotation, undergoing two revolutions for each full trap period as given by~\eqref{eq.Om}, sweeping the equator of the Bloch sphere (see Fig.~\ref{fig1}\textbf{b}).

We start with a relatively slow (compared to the condensate and reservoir GHz dynamical timescales~\cite{baryshev_prl}) $f = 2.5$ MHz counter-clockwise rotation ($l_1=1, l_2=-1$). The condensate pseudospin then follows the trap small axis adiabatically. For this, the trap diameter ($10.5$ $\mu$m) and the pump power ($ \approx 2 P_\text{th}$) are chosen to drive the condensate into the ground state of the trap, avoiding condensation into higher order modes~\cite{Topfer_PRB2020}. The time-integrated (averaged) measurement of total cavity emission, proportional to the total condensate density distribution, 
reveals a cylindrically symmetric profile (see Supplementary Information Section S1). Whereas resolving the vertical (V) and horizontal (H) linear polarisations of the condensate emission (defined along the $x$ and $y$ direction respectively) we observe instead elliptically-shaped spatial density distributions, as shown in Figs.~\ref{fig1}\textbf{c,d}. The ellipticity (ratio of the  big and small axes) of the rotating condensate is 1.2. Notably, the vertically polarised emission has a small axis along $y$-axis and the horizontal emission along the $x$-axis ~\cite{gnusov_prapl}. This indicates that the trap is rotating the condensate and its pseudospin follows in-step with the trap small axis.

\begin{figure}[t]
    \centering
    \includegraphics[width=0.8\columnwidth]{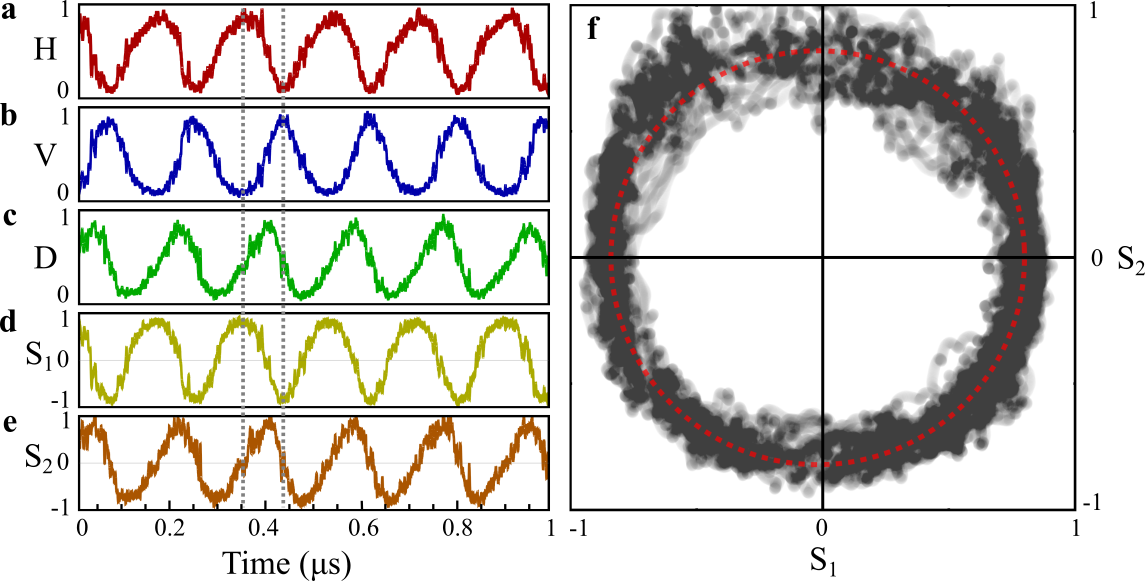}
    \caption{\textbf{Measured rotation of polariton condensate polarisation (spin) under slow stirring.} \textbf{a} Horizontal H, \textbf{b} vertical V and \textbf{c} diagonal D polarisation projections of the condensate emission. The ratio of excitation lasers intensities is 10\%. Time-resolved \textbf{d}~$S_1$ and \textbf{e}~$S_2$ polarisation components of the condensate emission. \textbf{f} Trajectory of the condensate spinor on the equator of Poincar\'{e} sphere. The red dashed line is to guide the eye. }
    \label{fig2}
\end{figure}

To resolve the linear polarisation rotation in time, we split the condensate emission into its polarisation components with a polarising beamsplitter (PBS) and detect the intensity of H and V polarisations with photomultiplier tube detectors (the bandwidth is $\approx 500$ MHz). 
Figures~\ref{fig2}\textbf{a,b} show the normalised readings for synchronously recorded sinusoidal oscillations in time for the H and V polarisations with $\pi$ phase difference. The depth of the signal amplitude modulation is close to 100\%, that underlines the efficiency of our trap modulation technique. By sampling a part of the condensate emission before the polarising beamsplitter, we simultaneously detect the diagonal (D) polarisation component~\cite{gnusov_prb} (which, along with the antidiagonal polarisation, corresponds to the $S_2$ pseudospin component). As expected, the D polarisation also demontrates clear sinusoidal oscillations with a $\frac{\pi}{2}$ phase shift with respect to the H and V polarisations, as shown in Fig.~\ref{fig2}\textbf{c}. The observed slight deviation from the perfect sinusoidal oscillations can be attributed to small polarisation pinning from cavity disorder making the condensate pseudospin lean towards a specific polarisation~\cite{kasprzak_bose-einstein_2006, Read_PRB2009, gnusov_prb,gnusov_prapl}.

The corresponding oscillating $S_1$ and $S_2$  components of the polariton condensate are depicted in Figs.~\ref{fig2}\textbf{d,e}. A more vivid picture of the pseudospin rotation is obtained by plotting $S_1$ and $S_2$ together revealing a circle on the equatorial plane of the Poincar\'{e} sphere with an average radius of $\approx 0.85$. The radius manifests the degree of linear polarisation in the system and its value below unity $<1$ implies some depolarisation present in the system. Here, it can be associated with exciton reservoir induced depolarisation and spin imbalance, and dynamical fluctuations in the condensate pseudospin~\cite{Ryzhov_PRR2020, baryshev_prl,gnusov_prb}. 
However, despite the noise, the achieved polarisation modulation is stable within the whole excitation duration, with the condensate traveling through all of the linear polarisation states. We stress that by either reversing the detuning of the two superposed incident pump beams, or their angular momenta, we can reverse the rotation of the trap~\cite{rotatingbucket} and the pseudospin (see Supplementary Information Section S2).

\begin{figure}[t]
    \centering
    \includegraphics[width=1\columnwidth]{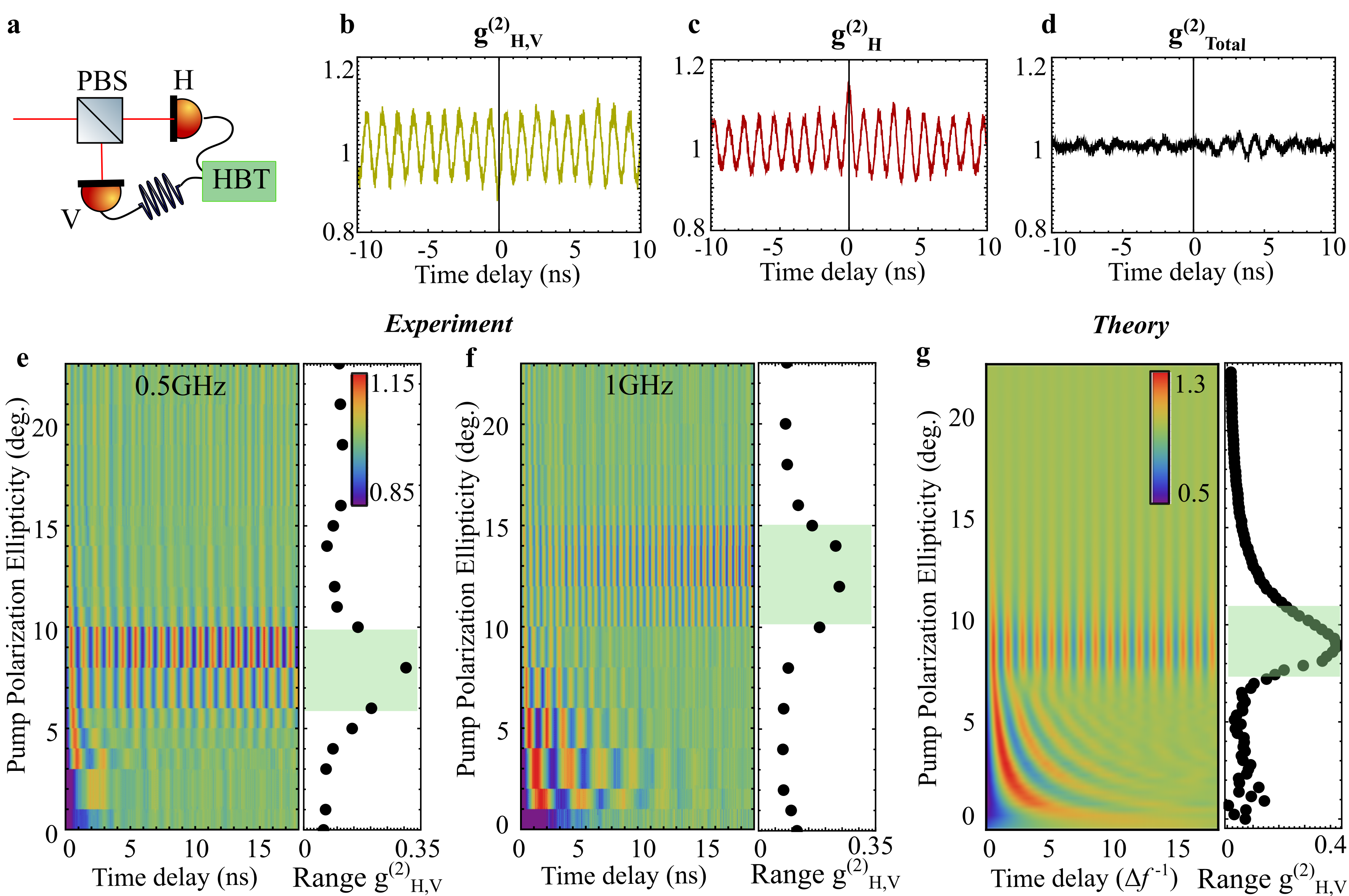}
    \caption{\textbf{GHz driven precession of the condensate pseudospin appearing in resonance with internal Larmor precession.} \textbf{a} The schematic of the detection part. The condensate emission is split on a PBS, and the H and V polarisations are correlated in HBT interferometer either in cross- or auto- correlation setting. Measured line profiles of the second-order H-V cross-correlation  $g^{(2)}_{H,V}$ \textbf{b}, H-H auto-correlation $g^{(2)}_{H,H}$ \textbf{c}, and total intensity auto-correlation $g^{(2)}_\text{Total}$ \textbf{d} functions versus delay at $8^\circ$ pump polarisation ellipticity. Measured H-V cross-correlation maps for~\textbf{e} $f = 0.5$ GHz and~\textbf{f} $f = 1$ GHz trap rotation as a function of delay and pump polarisation ellipticity. \textbf{g} Simulated $g^{(2)}_{H,V}$ of the condensate emission versus the time delay and pump polarisation ellipticity. The black point graphs on ~\textbf{e}, ~\textbf{f} and ~\textbf{g} respectively demonstrate the range (maximum minus minimum value) of the cross-correlation function $g^{(2)}_{H,V}$ in the 2 ns vicinity of 15~ns time delay as a function of pump ellipticity for a corresponding colormap image.
    }
    \label{fig3}
\end{figure}

In order to probe the pseudospin dynamics under faster GHz rotation frequencies wherein the trap rotates at a rate comparable to the condensate characteristic energy scales (e.g., the exciton-exciton interaction strength in GaAs cavities is around $\approx 2.5$ GHz~\cite{Kuriakose_NatPho2022}), we use the same experimental setup as in~\cite{baryshev_prl}. We employ a Hanbury Brown and Twiss (HBT) interferometer coupled with time-correlated single photon counting technique to measure the second-order cross-correlation function $g^{(2)}_{\mu,\nu}(\tau)$ between two signals $\mu$ and $\nu$---which can be different polarisation components of the emitted cavity light---as a function of time delay $\tau$ between them,
\begin{align}\label{eq.g2}
g_{\mu,\nu}^{(2)}(\tau)= \frac{\langle{a_{\mu}^{\dag}(t)}{a_{\nu}^{\dag}(t+\tau)}{a_{\nu}(t+\tau)}{a_{\mu}(t)}\rangle}{\langle{a_{\mu}^{\dag}(t)}{a_{\mu}(t)}\rangle\langle{a_{\nu}^{\dag}(t+\tau)}{a_{\nu}(t+\tau)}\rangle}.
\end{align}
Here, $a_{\mu}^\dagger$ and $a_{\mu}$ are photon creation and annihilation operators of the emitted cavity light for a given polarisation $\mu$. We point out that we have chosen this technique instead of measuring the first-order correlation function of the emission which is limited by the shorter global phase coherence of the condensate~\cite{coher_revivals}. 
Using a polarising beamsplitter (as before) we measure the photon coincidence statistics between two orthogonal polarisation projections of the emission H,V (see Fig.~\ref{fig3}\textbf{a}).

Interestingly, the driven spin precession under linearly polarised excitation holds until $f \approx$ 150 MHz. At this stage the adiabatic character of the rotation at higher frequencies is no longer valid, since the rotation is approaching the spin coherence time for the stationary condensate~\cite{baryshev_prl} and internal condensate spin dynamics starts to play a role. Therefore, setting the stirring frequency f= 0.5 GHz we do not observe the expected rotation of the condensate pseudospin under linearly polarised excitation (see Fig.~\ref{fig3}\textbf{e} at 0$^{\circ}$). This implies a deviation from the adiabatic condition with the pseudospin unable to keep up with the trap rotation. The spin dynamics is now found to resemble that of a stationary optical trap with anti-bunching at zero time delay corresponding to stochastic spin flips on the equatorial plane of the Poincar\'{e} sphere~\cite{baryshev_prl}.

The behaviour of the condensate pseudospin is drastically changed when we apply finite polarisation ellipticity to the pump lasers by inserting a quarter waveplate (QWP) into the excitation path. The 0$^{\circ}$ of the QWP fast axis  corresponds to linearly polarised excitation, and plus and minus 45$^{\circ}$ to right and left circularly polarised ones respectively. For the $f=0.5$ GHz and 8$^{\circ}$ of the QWP (elliptically polarised excitation) we observe a strong revival in the pseudospin oscillations seen in the H-V cross-correlation function, as visible from  Fig.~\ref{fig3}\textbf{e} and corresponding line profile in Fig.~\ref{fig3}\textbf{b}. The frequency of these oscillations is 1 GHz [equal to doubled trap rotation frequency $f$ described in eq.~\eqref{eq.Om}]. The $g_{H,V}^{(2)}(0)<1$ at zero time delay manifests the anti-correlation between H and V polarisation components. In contrast, the auto-correlation function of, e.g., the horizontal polarisation $g_{H,H}^{(2)}(0)$ depicted in Fig.~\ref{fig3}\textbf{f} possesses a local maximum at zero time delay as well as sine modulation, which is expected for the photon statistics of the light source with rotating polarisation. Figure~\ref{fig3}\textbf{d} shows the 
 intensity correlation function of the total emitted light $g^{(2)}_\text{Total}$ without polarisation filtering. Notably, it does not depict any oscillations, which evidences that the oscillations observed in $g^{(2)}_{H,V}$ and $g^{(2)}_{H,H}$ cannot be attributed to intensity modulation in the condensate but rather stable precession of the pseudospin on the equator of the Poincar\'{e} sphere. The fact that $g^{(2)}_\text{Total} \approx 1$ everywhere also implies that the condensate state is highly coherent. 

Remarkably, the observed pseudospin oscillations are extremely stable and persistent (see Supplementary Information section S3). We find their decay time to be ($174 \pm 10$) ns, which is approximately 20 times longer than the record polariton spin coherence time reported recently~\cite{baryshev_prl}. Utilising the analogy with the damped pendulum, we deduce the quality factor $Q$ of spin oscillations of 546. In our experiment, only the mutual frequency stability of the utilised pump lasers limits the experimentally obtained spin coherence time. However, we believe that the use of alternative techniques with improved stability of frequency detuning could potentially increase the spin coherence time and quality factor even more.

On the inset of Fig.~\ref{fig3}\textbf{e} we plot the range of the H-V cross correlation amplitudes (the difference of maximum and minimum $g^{(2)}_{H,V}$) in the 2 ns vicinity of 15 ns time delay versus the pump ellipticity. This curve highlights the resonant nature of the observed driven precession, with the resonance appearing near the 8$^{\circ}$. The resonant-like behaviour of the spin coherence in the polariton system can be understood as a driven-dissipative quantum fluidic analogue of the nuclear magnetic resonance effect~\cite{nmr_rabi}, also reported in optically driven dilute atomic gases~\cite{opt_driv_PhysRevLett.6.280} displaying persistent spin precession without decay. In our system, an additional effective out-of-plane magnetic field appears due to the strong spin-anisotropic polariton interactions of the spin up and down polaritons populations when they are imbalanced by an elliptically polarised pump~\cite{Renucci_PRB2005}. This effective out-of-plane magnetic field gives rise to the self-induced Larmor precession, which also rotates the pseudospin along the equator of the Poincar\'{e} sphere~\cite{shelykh_spin_dyn_2004, larmor_polariton2006, baryshev_prl, coher_revivals}. Formally, the total effective magnetic field under these conditions can be written
\begin{equation} \label{eq.Om2}
\boldsymbol{\Omega}_\text{Total} = \begin{pmatrix} \Omega_\parallel \cos{[2  \omega t]} \\ \Omega_\parallel \sin{[2 \omega t]} \\ \Omega_\perp
\end{pmatrix},
\end{equation}
where the out-of-plane component is written $\Omega_\perp = \Omega_+ - \Omega_-$ and $\Omega_\pm$ are the different blueshifts experienced by the condensate polaritons when the pump is elliptically polarised (see Methods for theoretical model). For simplicity, we can say that the pump is parameterised into $P_\pm$ components denoting the intensity of right and left-hand circularly polarised incident light and that approximately $\Omega_\pm \propto P_\pm$. This implies that at small angles
\begin{equation}\label{eq.Omp2}
\Omega_\perp \propto P_0 \sin{(2\Theta)} \approx 2 P_0 \Theta
\end{equation}
where $P_0$ is the total pump power and $\Theta$ quantifies the pump ellipticity (i.e., the QWP angle). When $\Omega_\perp = \omega$ the trap rotation is in resonance with the self-induced Larmor precession and the pseudospin starts rotating with renewed stability. 
We have performed simulations using a generalized Gross-Pitevskii simulations (see Methods) accounting for the effective magnetic field~\eqref{eq.Om2} and reproduced the resonance with good accuracy (see Fig.~\ref{fig3}\textbf{g}).

Under the same experimental conditions but with the doubled  trap rotation frequency, $f = 1$ GHz, the same resonant effect is observed but now at approximately double the value of pump polarisation ellipticity $\approx 13^{\circ}$ with respect to the 0.5 GHz rotation (see Fig.~\ref{fig3}\textbf{f}). This further confirms our interpretation that the revival of the spin oscillations is due to a resonance between the condensate induced effective magnetic field $\Omega_\perp$~\eqref{eq.Omp2} and the rotating trap effective field~\eqref{eq.Om}.

\section*{Discussion}
We have demonstrated persistent GHz precession of a macroscopic pesudospin belonging to a spinor microcavity exciton-polariton condensate under incoherent continuous time-periodic optical driving. Power driven Bose-Einstein condensation of two-component spinor exciton-polaritons takes place in the optical trap with a macroscopic pseudospin pinned along the trap minor axis. As the trap rotates the pseudospin follows suit, thus demonstrating time-modulated control over the internal state of a polariton condensate. 

Moreover, we uncover experimentally the resonant effect in driven-dissipative quantum fluids which can only be attributed to the strong spin-anisotropic interactions of polaritons. Namely, when the drive beam is elliptically polarised the condensate experiences and additional effective out-of-plane magnetic field which competes with the trap-induced in-plane field. We observe a sharp resonant feature reminiscent of the nuclear magnetic resonant effect but, this time, in a quantum fluid of light. The broken axial symmetry of the rotating exciton reservoir in conjunction with elliptically polarised excitation effectively mimics the roles of a NMR transverse rf field and a holding field, respectively. Our findings are corroborated through mean field calculations using the stochastic generalized spinor Gross-Pitaevskii model. At resonance, the condensate pseudospin precession features striking stability and record 174 ns spin coherence time, more than an order magnitude higher than previous reports with a static optical trap~\cite{baryshev_prl}.

 The 20-fold increase in spin coherence opens great perspectives for novel spinoptronic devices and quantum computing~\cite{Kavokin2022} based on polariton condensates. Our findings will potentially allow coherent control of the condensate spin state at longer timescales for the realisation of polariton quantum bits. Moreover, recent developments in atomically thin materials~\cite{Lundt_NatNano2019} and cavities embedded with highly birefringent liquid crystals~\cite{Rechcinska_Science2019} have demonstrated strong synthetic spin-orbit-coupling which, alongside advances in structuring large-scale lattices of polariton condensates~\cite{Alyatkin_NatComm2021}, could be implemented in our scheme to explore spatially structured spin-orbit-coupled polariton condensates under time-periodic effective magnetic fields. 
 
Unlike the conventional spin switches realised in magnets, polaritonics, or VCSELs, our results show that the condensate pseudospin sweeps through all linear polarisation states on the equator of the Poincar\'{e} sphere. We also achieve GHz polarisation modulation, which is not feasible for conventional crystal polarisation modulators.
Moreover, we have recently demonstrated the possibility to spin up vortices in rotating polariton condensates~\cite{rotatingbucket} that, in conjunction with our effective magnetic fields achieved here, could pave the way to realise skyrmion and meron topological textures in nonequilibrium polariton condensates~\cite{Yulin_PRB2020}. The advantage of being able to optically access such exotic states through the polariton emission is in contrast to the more difficult experiments on equilibrium ultracold atom condensates~\cite{Matthews_PRL1999, Schweikhard_PRL2004, Xu_PRL2011}.

\section*{Methods}\label{sec5}

\bmhead{Modeling}
The dynamics of the spinor polariton condensate order parameter is modeled through a set of stochastic driven-dissipative Gross-Pitaevskii equations (Langevin type equations) coupled to spin-polarised rate equations describing excitonic reservoirs $X_\pm$ feeding the two spin components $\psi_\pm$ of the trap ground state condensate~\cite{Read_PRB2009, Wouters_PRB2009}.
\begin{subequations} \label{eq.orig}
\begin{align} \label{eq.1a}
i & \frac{ d\psi_\sigma}{dt}  =   \Big[ \Omega_\sigma+ \frac{i}{2} \left( R X_\sigma - \Gamma \right) \Big] \psi_\sigma + \Omega_\parallel \psi_{-\sigma} e^{-\sigma i 2 \omega t} + \theta_\sigma(t), \\ 
& \frac{ dX_\sigma}{dt}  =   - \left( \Gamma_R + R\lvert\psi_\sigma \rvert^2  \right)X_\sigma + \Gamma_s(X_{-\sigma} - X_\sigma) + P_\sigma,\\ \label{eq.1c}
& \Omega_\sigma = \alpha \lvert\psi_\sigma \rvert^2  + G\left(X_\sigma + \frac{P_\sigma}{W}\right).
\end{align}
\end{subequations}
Here, $\sigma \in \{+, -\}$ are the two spin indices, $\alpha$ and $G$ denote the same-spin polariton-polariton and polariton-reservoir interaction strengths, $R$ is the rate of stimulated spin-conserving scattering of reservoir excitons into the condensate, and $\Gamma$ is the polariton decay rate, $\Gamma_R$ and $\Gamma_s$ describe the decay rate and spin relaxation~\cite{Maialle_PRB1993} of reservoir excitons. We neglect opposite-spin interaction as it is usually much weaker than same-spin interactions, and we also neglect cross-spin scattering into the condensate from the reservoir as it did not qualitatively affect our results. The reduced symmetry of the elliptical trap shape, alongside TE-TM splitting from the cavity mirrors, causes fine-structure splitting in the polariton condensate described by an effective in-plane magnetic field $\Omega_{\parallel}$ along the trap minor axis~\cite{gnusov_prapl}. Consequently, the effective field rotates at a frequency $2 \omega$.

Without gain-decay and nonlinearities Eq.~\eqref{eq.1a} becomes the celebrated Rabi model describing a two-level atom interacting with a monochromatic electromagnetic field of frequency $2 \omega$ in the rotating wave approximation. In this sense, the trap birefringence $\Omega_\parallel$ plays the role of an effective dipole moment. Another famous phenomenon corresponding to our model and experiment is nuclear magnetic resonance wherein a two-level spin (i.e., a spin 1/2 particle) is subject to a time-dependent magnetic field $\boldsymbol{B} = (B_x \cos{(2 \omega t)}, B_y\sin{(2\omega t)}, B_z)^\text{T}$. In our case, the time-dependent in-plane magnetic field corresponds to the oscillating birefringence from our spinning non-axisymmetric trap $B_{x,y} = \hbar \Omega_\parallel/\mu$ where $\mu$ is the Bohr magneton. The out-of-plane magnetic field in our experiment comes from the spin-anisotropic interactions of between polaritons $B_ z = \hbar \Omega_\perp/\mu = \hbar(\Omega_+ - \Omega_-)/\mu$ given by Eq.~\eqref{eq.1c} which is non-zero when the pump beam is elliptically polarised $P_+ \neq P_-$. Notice that if the beam ellipticity is reversed then the sign of the effective field $\Omega_\perp$ flips.

The white complex-valued noise $\theta_\sigma(t)$ is determined by the correlators,
$\langle \theta_\sigma(t) \theta_{\sigma'}(t') \rangle = 0$ and $\langle \theta_\sigma(t) \theta^*_{\sigma'}(t') \rangle  = \eta \delta_{\sigma \sigma'} \delta(t-t')$ where $\eta$ is the noise amplitude (typically proportional to the scattering rate of polaritons into the condensate). The incoherent drive $P_\sigma$ for the exciton reservoir represents a background population of higher-momentum excitons (so called, {\it inactive reservoir} consisting of excitons which do not scatter into the condensate). Assuming the non-radiative lifetimes of these excitons to be very long compared to the polariton lifetime, we can write $P_+ = P_0(W \cos^2{(\theta)} + \Gamma_s)/(W+2\Gamma_s)$ and $P_- = P_0(W \sin^2{(\theta)} + \Gamma_s)/(W+2\Gamma_s)$~\cite{baryshev_prl}. Here, $P_0$ is the total power of the pump, $W$ is a phenomenological spin-conserving redistribution rate between high-momentum and low-momentum excitons. The calculated $g^{(2)}_{H,v}(\tau)$ is obtained by replacing the operators $a^\dagger_\mu$ and $a_\mu$ with the mean fields $\psi^*_\mu$ and $\psi_\mu$ obtained by solving Eq.~\eqref{eq.1a} in time.

In our simulation presented in Fig.~\ref{fig3}g we have fitted the experiment with the following parameters in units of $\omega$: $\Gamma = 100$; $\Omega_\parallel = 0.16$; $R = 6$; $G=0.2$; $\alpha = 1$; $W = \Gamma/4$; $\Gamma_R = \Gamma_s = \Gamma/2$; $P_0 = 2.2 \Gamma \Gamma_R/R$; $\sqrt{\eta} = 0.01$.

\section*{Acknowledgments}
H.S. acknowledges the Icelandic Research Fund (Rannis), grant No. 239552-051.

\section*{Contributions}

I.G., S.A., S.B. and  K.S. performed the experiments and analysed the data.  J.T. developed the software for the data acquisition.  H.S. developed the
theory and carried out numerical simulations. I.G., S.B. and P.G.L. designed the experiment. P.G.L. designed and led the research project. All the authors participated in the writing of the manuscript.

 \section*{Competing interests.} 

Authors declare that they have no competing interests.    

 \section*{Data and Materials Availability.} 
 
All data supporting this study are openly available from the University of Southampton repository [link is to be provided].


\setcounter{equation}{0}
\setcounter{figure}{0}
\setcounter{section}{0}
\renewcommand{\theequation}{S\arabic{equation}}
\renewcommand{\thefigure}{S\arabic{figure}}
\renewcommand{\thesection}{S\arabic{section}}

\newpage
\vspace{1cm}
\begin{center}
\Large \textbf{Supplementary Information}
\end{center}

\section{Pump profile and condensate Intensity distribution.}

\begin{figure}[h]
    \centering
    \includegraphics[width=0.6\columnwidth]{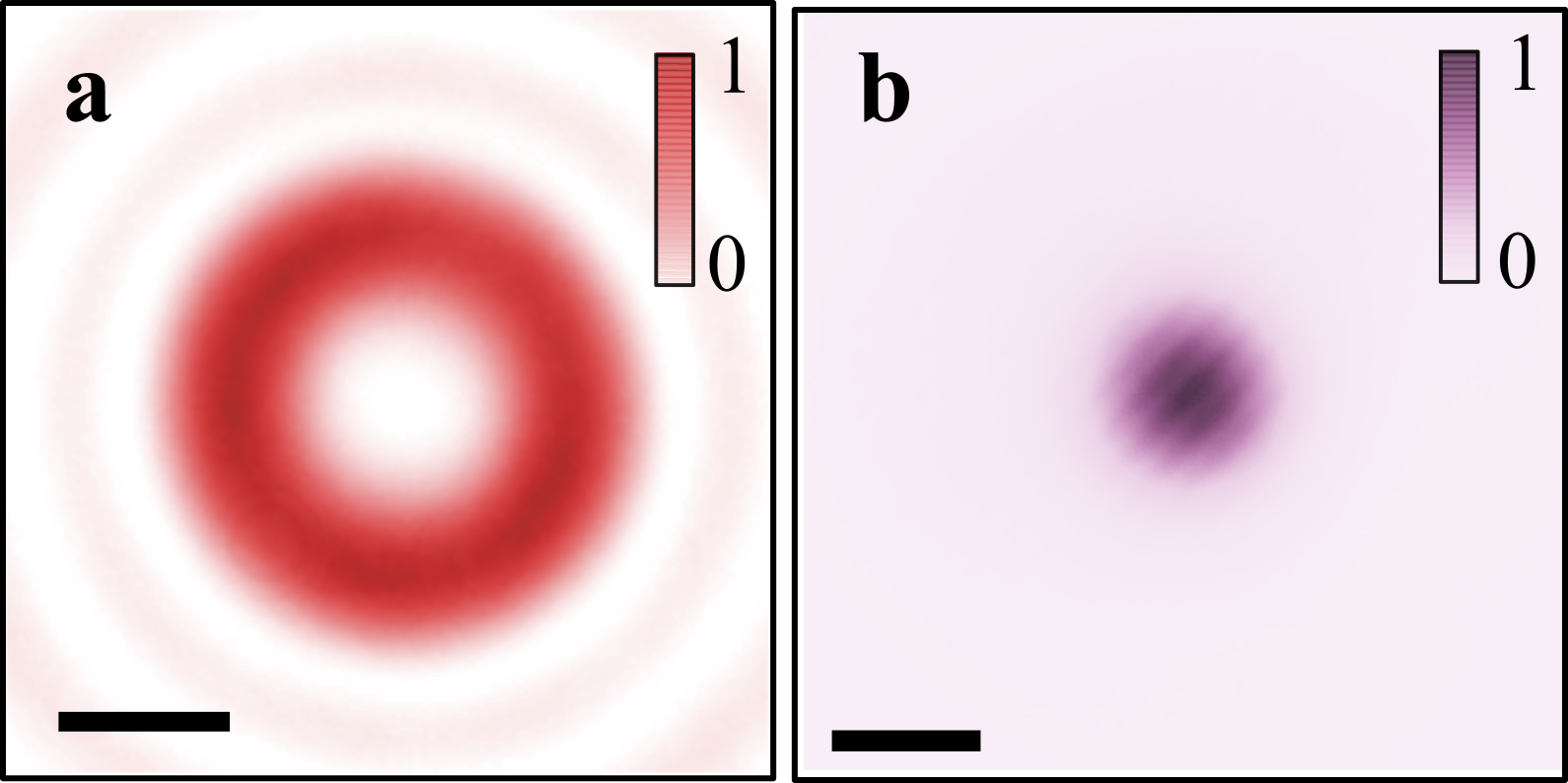}
    \caption{\textbf{a} The normalised time-integrated intensity of one of the excitation lasers. \textbf{b} The normalised time-integrated intensity distribution of the polariton condensate at 2$P_{th}$. The black scale bar corresponds to $5\mu m$.
    }
    \label{pump}
\end{figure}

The polariton condensate in our experiments is created by the non-resonant rotating laser pattern formed by the beating note of two frequency-stabilized detuned lasers. Each of the laser beams is shaped into the "perfect vortex" - the ring with controlled radius and optical angular momenta. The intensity distribution of one of the excitation lasers is depicted in Figure~\ref{pump}\textbf{a}. The intensity profile of the second laser is identical. We note, that the intensity distribution is uniform around the ring circumference. For MHz trap rotation frequencies we intensity imbalanced the two interfering beams so that the intensity of one of the excitation lasers is 10\% with respect to that of the other to achieve a clear elliptical excitation pattern focused on the cavity plane. However, for the GHz rotation speed, we increased this ratio to 20\% in order to compensate for finite exciton lifetime which leads to effective smearing of trap profile[32].

Overlapping the two laser beams in space and time, we achieve the rotation of the excitation profile for polaritons. The resultant time-integrated polariton condensate emission measured at 2$P_{th}$ is depicted in Figure~\ref{pump}\textbf{b}. We note, that such time-integrated (over 10 ms)  profile is cylindrically symmetric, whereas the polarisation resolved intensity profiles appear to be elliptical (see Figure1 \textbf{a,b} in the main text). 

\section{The control over the polarisation rotation direction.}

 The rotation frequency of the optical trap used in the experiments is defined by the equation~\eqref{rotfreq}, where $f_1,f_2$ and $l_1,l_2$ are the frequencies and optical angular momenta of two excitation lasers respectively. The direction of the trap rotation is dictated by the sign of the $f$, and the positive and negative values correspond to the counter- and clockwise rotations respectively. Thus, by changing the relative values (ergo, the sign of their difference) of either the excitation lasers frequencies or their OAMs one can control the trap rotation direction as well as the condensate spin defined by the trap orientation (i.e., the angle with respect to horizontal axis). 

\begin{equation} \label{rotfreq}
 f=\frac{\omega}{2\pi} = \frac{\Delta f}{\Delta l}=\frac{f_1-f_2}{l_1-l_2}.
 \end{equation}

\begin{figure}[h]
    \centering
    \includegraphics[width=0.8\columnwidth]{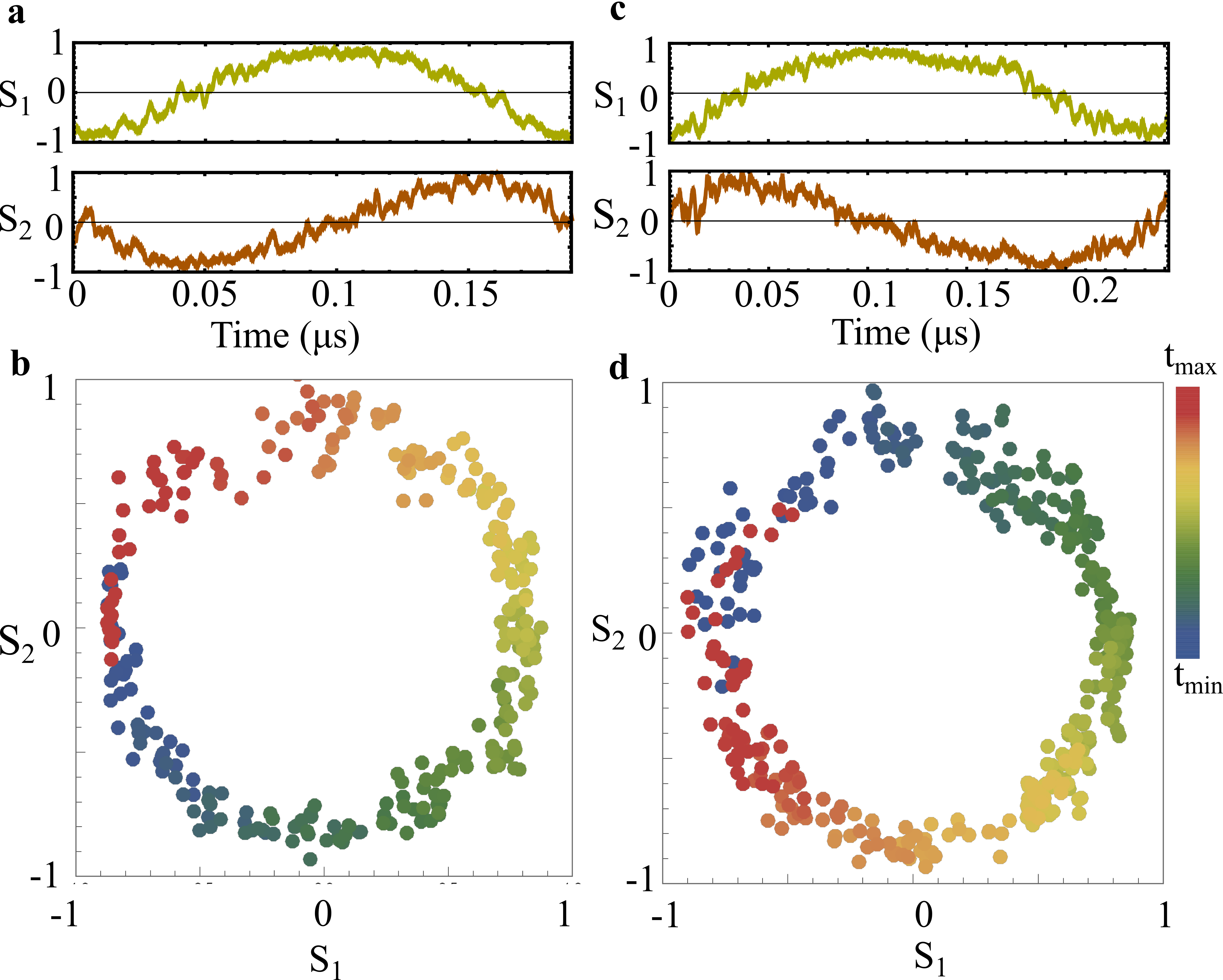}
    \caption{\textbf{Control on the direction of the driven spin precession.} \textbf{a} $S_1$ and $S_2$ polarisation components of the polariton condensate emission versus time. Here, the condensate forms in the rotating trap with the $\Delta f \approx 5$ MHz and $l_1 = 1,l_2 = -1$. \textbf{b} The corresponding  Stokes vector trajectory on the equator of the Poincare sphere. \textbf{c} Time-resolved $S_1$ and $S_2$ polarisation components of the condensate emission for the  $\Delta f \approx 5$ MHz  and $l_1 = -1,l_2 = 1$, and the corresponding trajectory~\textbf{d} in the linear polarisation plane of the Poincare sphere. The color scale in panels ~\textbf{b,d} represents the time passing.   
    }
    \label{rotdir}
\end{figure}

To probe the change in the polarisation rotation direction, we perform the same experiment as for the Figure 2 in the main text. At ($\Delta f \approx 5$ MHz) rotation frequency we first set  $l_1 = 1,l_2 = -1$,  and the resultant time series for the linear polarisation Stokes components $S_1$ and $S_2$ of the condensate emission are presented in Figure~\ref{rotdir}\textbf{a}. Plotting the obtained Stokes components on the equator of the Poincare sphere (see Figure~\ref{rotdir}\textbf{b}) and tracking the trajectory of the linear polarisation, we obtain the counter-clockwise rotation of the condensate spin. On the contrary, flipping the OAMs of the interfering beams ($l_1 = -1,l_2 = 1$) at the same sign of frequency detuning $\Delta f \approx 5 MHz$ results in the clock-wise rotation of the condensate spin as depicted in Figures~\ref{rotdir}\textbf{b,d}. Thus, with our technique for the polarisation rotation (driven spin precession) we as well gain the control over the direction of the induced rotation. 

\section{H-V cross-correlation at a longer time delays.}

\begin{figure}[h]
    \centering
    \includegraphics[width=1\columnwidth]{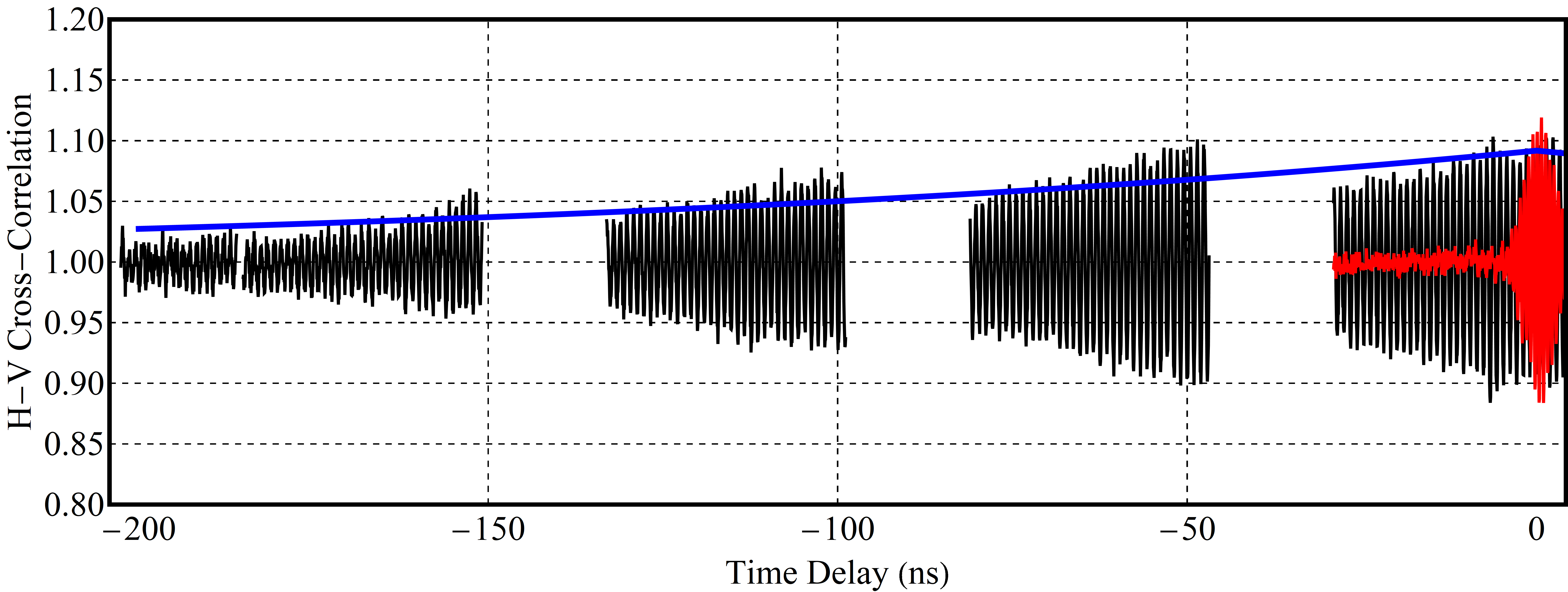}
    \caption{ \textbf{Measured $g^{(2)}_{H,V}$ of the condensate emission at big time delay.} The H-V cross correlation $g^{(2)}_{H,V}$ for the condensate in the rotating optical trap (black curve) and stationary ring optical trap excited with single laser (red curve). The blank spaces in the black graph appear due to incremental addition of the optical delay and limited measurement window.  The blue solid line is the fit of the experimental data envelope with $Ae^{-t/\tau}$.
    }
    \label{longdelay}
\end{figure}

The driven spin precession is persistent in the whole 30 ns HBT measurement window (see Figure 3 in the main text). In order to probe the temporal stability of the oscillations at bigger time scales we measure the cross-correlation of the H and V polarisation components at a longer time delay. To do so we add additional optical delay to one of the correlating signal path. We incrementally add few 10 meter optical fibers (each corresponds to 52 ns delay)  in one arm of the HBT interferometer and measure the $g^{(2)}_{H,V}$ for the increased time delay. We add up to 31 m ($\approx 162$ ns delay) of optical fiber as well as some electronic delay available with the HBT apparatus (so the maximum time delay achieved is $\approx 200$ ns). The obtained results are presented in Figure~\ref{longdelay}, or the comparison, we also plot in the same graph the self-induced Larmor precession of the condensate under elliptically polarised excitation (see the  red curve in Figure~\ref{longdelay}). The amplitude of the driven oscillations is dropped twice at approximately 150 ns, even though the precession is still distinguishable up to 200 ns. We fit the envelop of the oscillations presented in in Figure~\ref{longdelay} with the decay exponent ($Ae^{-t/\tau}$), where $A$ is an amplitude of the oscillations at zero time delay, t is time, $\tau$ is spin coherence (decay) time. From the fit we retrieve the value of $\tau = (174\pm10)$ns. Further, we estimate the quality factor $Q$ of the driven precession $Q= \tau\omega/2$, where $\omega = 1$ GHz is precession frequency, to be 546. 

We underline here, that the decrease in $g^{(2)}_{H,V}$ at big time delay appears due to not perfect mutual stability of two excitation lasers. Frequency jitter during the acquisition time results into the effective tiny change of the stirring frequency, which smears the $g^{(2)}_{H,V}$ at a bigger time delays. We note, that the measurement of one $g^{(2)}_{H,V}$ curve in our experiments takes at least 20 minutes, which implies the integration over $6*10^6$ realisations (2$\mu$s pulses) of the condensate. However, we stress that the spin precession is present in the condensate for the whole $\mu s$excitation pulse (see Figure 2 in the main text).

\end{document}